\documentclass[12pt]{article}
\usepackage{epsfig,overcite}
\newcommand{\be}{\begin{equation}}
\newcommand{\ee}{\end{equation}}
\newcommand{\bea}{\begin{eqnarray}}
\newcommand{\eea}{\end{eqnarray}}
\newcommand{\beas}{\begin{eqnarray*}}
\newcommand{\eeas}{\end{eqnarray*}}
\newcommand{\ba}{\begin{array}}
\newcommand{\ea}{\end{array}}

\newcommand{\eps}{\epsilon}
\newcommand{\rf}[1] {(\ref{#1})}
\def\C{{\rm\kern.24em \vrule width.02em height1.4ex depth-.05ex \kern-.26em
C}}
\def\R{{\rm I\kern-.20em R}}
\def\Z{{\rm\kern.26em \vrule width.02em height0.5ex depth0ex \kern.04em
\vrule  width.02em height1.47ex depth-1ex \kern-.34em Z}}

\title{Predicting rogue waves in random oceanic sea states}

\author{A. L. Islas  and C.M. Schober\footnote{Corresponding
author: Phone: 407-823-0147; Fax: 407-823-6253;
email: cschober@mail.ucf.edu}\\
\\Dept. of Mathematics, University of Central Florida,\\
Orlando, Florida 32816}

\date{}

\begin{document}

\maketitle
\begin{quote}
Using the inverse spectral theory of the 
nonlinear Schr\"odinger (NLS) equation
we correlate the development of rogue waves
in oceanic sea states characterized by the JONSWAP spectrum 
with the proximity to homoclinic solutions of the 
NLS equation. We find 
in numerical simulations of the NLS equation that 
rogue waves develop 
for JONSWAP initial data that is ``near'' NLS homoclinic data,
while rogue waves do not occur for JONSWAP data that is ``far'' from 
NLS homoclinic data.
We show the nonlinear spectral decomposition  provides a simple criterium
for predicting the occurrence and strength of rogue waves
(PACS: 92.10.Hm, 47.20.Ky, 47.35+i).
\end{quote}

\noindent {\bf Introduction}
Rogue waves are rare, large amplitude waves
whose 
heights exceed 2.2 times the significant wave height of the background 
sea.
One of the proposed mechanisms for the development of
rogue waves in deep water is 
nonlinear focusing due to the Benjamin-Feir (BF) instability.$^{1,2}\;$
The BF instability is a modulational instability in which a uniform 
train of surface gravity waves is unstable to a weak 
amplitude perturbation.
The BF instability is described  approximately by the focusing 
nonlinear Schr\"odinger (NLS)  equation  $^{3}\;$
\be
iu_{t} + u_{xx}  + 2|u|^2u = 0,
\label{NLS}
\ee
and in the simplest
setting homoclinic orbits of the 
unstable  Stokes solution of the NLS equation have
been used for modeling rogue waves\cite{oz2000,cs02}.
Homoclinic solutions of the NLS equation, obtained
when two or more unstable modes are present, can be phase modulated
to provide striking examples of wave amplification where 
the amplification is due to both the BF instability 
and the additional phase modulation\cite{cs02}.

The NLS equation is the leading order equation in a hierarchy of
envelope equations and is derived from the 
full water wave equations under the assumption of a narrow 
$\cal O(\epsilon)$ banded spectrum.
This bandwidth constraint  limits the applicability
of the NLS equation in 2D as it results in energy leakage to
high wave number modes\cite{td1996}.
The broader bandwidth NLS (BBNLS) equation, obtained by
assuming the bandwidth is $\cal O(\sqrt\epsilon)$ and
by retaining higher order terms in the asymptotic  
expansion for the surface wave displacement,
has been successful in reducing the energy leakage\cite{td1996}.
An alternate approach is to ``enhance'' 
the  NLS equation with exact linear dispersion, 
whereby the equation
has improved bandwidth resolution
and stability properties\cite{tkd2000}.
All of these higher order equations, whether narrow or broader
bandwidth or ``enhanced'', may be viewed as perturbations of
the NLS equation.
Homoclinic orbits of the Stokes wave have been shown to
persist for the BBNLS equation.$^{5,8}\;$
This  persistence result suggests homoclinic solutions
of the NLS equation may be significant
in modeling rogue waves for random oceanic states.

Onorato {\it et al.}\cite{on2001a}
examined the generation of extreme waves for typical random 
oceanic sea states characterized by the Joint North Sea Wave Project
(JONSWAP)  power spectrum.
In numerical simulations of the NLS equation it was  found that 
rogue waves occur more often 
for large values of the Phillips parameter
$\alpha$ and the enhancement coefficient $\gamma$ 
in the JONSWAP spectrum. Even so,
they observed that large values of $\alpha$ and $\gamma$ do 
not guarantee the development of extreme waves.

In this paper we clarify the dependence of 
rogue wave events on the phases in the  
``random phase''  reconstruction of the surface elevation 
(see eqn. \rf{uinit}). 
We find that the  phase information is as important as
the amplitude and peakedness of the wave (governed by $\alpha$ and $\gamma$)
when determining the occurrence of rogue waves.
Random  oceanic sea states characterized by JONSWAP data
are not small perturbations 
of Stokes wave solutions. As a consequence,
it is difficult to investigate  the generation of
rogue waves in  more realistic sea states 
using a linear stability analysis (as in the 
Benjamin-Feir instability).
Our  approach is based on the NLS equation and its
inverse spectral theory, used to examine a nonlinear mode decomposition of  
JONSWAP type initial data. Such analysis allows us to
determine the nonlinear mode content of the data and the
proximity (measured in terms of a parameter $\delta$)
to instabilities and homoclinic solutions of the NLS equation.

Our main results are:
1)  JONSWAP data can be quite near data for homoclinic
orbits of complicated $N$-phase solutions.
For  fixed values  of $\alpha$ and $\gamma$ in the JONSWAP spectrum,
as the phases in the initial data are randomly varied, the
proximity $\delta$ to homoclinic data varies.
2) In several hundred simulations of the NLS, where the parameters 
and the phases in the JONSWAP initial data are varied, 
we find that rogue waves develop 
for JONSWAP data that is ``near'' NLS homoclinic data,
while rogue waves do not occur for JONSWAP data that is ``far'' from 
NLS homoclinic data.
Consequently, 
we find that the nonlinear spectral decomposition  provides 
a simple criterium, in terms of the proximity to homoclinic
solutions, for predicting the occurrence and strength of
rogue waves.
This is the first time homoclinic solutions have been correlated with 
rogue waves for realistic oceanic conditions.

\noindent {\bf Random oceanic sea states} 
To examine the generation of rogue waves in a random sea state,
we note that the surface elevation $\eta$ is related to $u$, the
solution of the NLS equation, by
\(\eta = \mbox{Re}\left\{iue^{ikx}\right\}/\sqrt{2}k\). Using
the Hilbert transform of $\eta$ and its associated analytical
signal, the initial condition for $u$
can be modeled as the random wave process
\be
\label{uinit}
u(x,0) = \sum_{n=1}^{N} C_n
\exp\left[i\left(k_{n-1} x - \phi_n\right)\right],
\ee
where $C_n$ 
is the amplitude of the $n$th component with 
 wave number 
$k_n = (n-1)k$, $k = 2\pi/L$,  and  random phase 
$\phi_n$, uniformly distributed on
the interval $(0,2\pi)$. The spectral amplitudes,
$C_n = -i\sqrt{2 S_n/L}$,  are obtained from  
the JONSWAP spectrum:\cite{on2001a}
\be
\label{jonswap}
S(f) = \frac{\alpha}{f^5}
\mbox{exp}\left[-\frac{5}{4}\left(\frac{f_0}{f}\right)^4\right]
\gamma^r,\qquad
r = \mbox{exp}\left[-\frac{1}{2}\left(\frac{f-f_0}{\sigma_0 f_0}\right)^2\right].
\ee
Here 
$f_0$ is the dominant frequency, determined by the wind speed at a 
specified height above the sea surface, 
$\sigma_0 = 0.07\; (0.9)$ for $f \leq f_0\; (f > f_0)$
and  $f_n = n/L$ is the wave frequency.
The parameter $\gamma$ is the peak-shape parameter; 
as $\gamma$ is increased, the spectrum becomes narrower about 
the dominant peak. 
For $\gamma > 1$ the wave spectra continues to evolve
through nonlinear wave-wave interactions 
even for very long times and distances.
It is in this sense that JONSWAP spectra describe developing sea states
rather than a fully developed sea.
The scale parameter
$\alpha$ is related to the amplitude and energy content of the wavefield.
Based on an ``Ursell number'', the ratio of
the nonlinear and dispersive terms of the NLS equation \rf{NLS} in dimensional 
form,  the NLS equation is considered to be applicable
 for $2<\gamma < 8$\cite{on2001a}.
Typical values of alpha are 
$.008<\alpha<.02$.

We examine a
nonlinear spectral decomposition of the JONSWAP initial data,
which takes into account the phase information $\phi_n$. This decomposition 
is based upon the inverse scattering theory
of the NLS equation,  a procedure 
for solving the initial value problem analogous
to Fourier methods for linear problems.
We find that we are able to predict the occurrence of rogue waves
 in terms of the 
proximity $\delta$ to distinguished points of the discrete spectrum.
We briefly recall elements of the nonlinear spectral theory of 
the NLS equation.

\noindent {\bf Floquet spectral theory}
The integrability of the NLS equation \rf{NLS} is related to
the following pair of linear systems (the so-called Lax pair)
\be
{\mathcal L}^{(x)}\phi = \left( \begin{array}{cc}
D + i\lambda & -u \\
u^* & D -i\lambda
\end{array} \right)
\left( \ba{c}
\phi_1\\ \phi_2
\ea \right) = 0,\quad
{\mathcal L}^{(t)}\phi = 0,
\label{lax1}
\ee
where $D$ denotes the derivative with respect to $x$,
 $\lambda$ is the spectral parameter 
and $\phi$ is the eigenfunction\cite{ref7}.
These systems have a common nontrivial solution $\phi(x,t;\lambda)$,
provided the potential $u(x,t)$ satisfies the NLS equation.
$\mathcal L^{(t)}$ is not specified explicitly as it is  not 
implemented  in our analysis.

The first step in solving the NLS using the inverse scattering
theory  is to determine  the spectrum
$\sigma(u) 
= \left\{ \lambda \in \C \ |\ {\mathcal L}^{(x)} \phi = 0, |\phi| 
\mbox{ bounded } \forall x\right\}$
of the associated linear operator ${\mathcal L}^{(x)}$, which 
is analogous to calculating the Fourier coefficients in Fourier theory.
For periodic boundary conditions, $u(x+L,t)=u(x,t)$, 
the spectrum of $u$ is expressed 
in terms of the transfer matrix $M(x+L;u,\lambda)$ across a period, 
where  $M(x;u,\lambda)$ is a 
fundamental solution matrix of the Lax pair \rf{lax1}. Introducing the 
Floquet discriminant 
$\Delta(u,\lambda) = \mbox{Trace}\left[ M(x+L;u,\lambda)\right]$, 
one obtains\cite{ref7}
\be
\sigma(u) = \left\{ \lambda \in \C \ | \ \Delta(u,\lambda) \in 
\R ,\ -2 \leq \Delta(u,\lambda) \leq 2 \right\} .    \label{spc}
\ee
The distinguished points of the periodic/antiperiodic spectrum,
where $\Delta(\lambda,u)=\pm 2$, are:
(a) simple points $\{\lambda_j^s \ | \ d\Delta/d\lambda \neq 0\}$
and (b) double points $\{\lambda_j^d \ | \
d\Delta/d\lambda = 0,\,d^2 \Delta/d\lambda^2 \neq 0\}$.
The Floquet discriminant functional $\Delta(u,\lambda)$ 
is invariant under the NLS flow and
encodes the infinite  family of constants of motion 
of the NLS (parametrized by the $\lambda_j^s$).

The Floquet spectrum \rf{spc} of a generic NLS potential consists
of the  entire
real  axis plus additional curves (called bands) of continuous spectrum 
which terminate at the simple points $\lambda_j^s$.
$N$-phase  solutions are those with a
finite number of bands of continuous spectrum. 
Double points arise when two simple points have coalesced
and their location is important.

Using the direct spectral transform, 
any initial condition  or solution of the NLS can be represented in 
terms of a set of nonlinear modes. The spatial
structure and dynamical
stability of these modes 
is determined by the order and location of the corresponding 
$\lambda_j$ as follows\cite{ref14}: 
(a) Simple points 
correspond to  stable active degrees of freedom.
(b) Double points label 
all additional potentially active degrees of freedom.
Real double points correspond to   stable
inactive (zero amplitude) modes.
Complex double points are associated with 
all the unstable active  modes 
and label the corresponding homoclinic orbits.

Figure 1 shows the spectrum of a typical unstable $N$-phase 
solution. 
\begin{figure}[htb]
\centerline{\psfig{figure=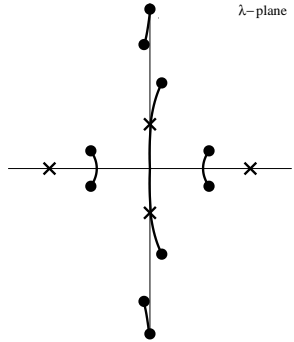,height=2.in}}
\caption{Spectrum of an unstable $N$-phase solution.}
\label{jons1}
\end{figure}
There are $N$ bands of spectrum 
determined by the $2N$ simple points $\lambda_j^s$.
The $2M$ complex double points
$\lambda_j^d$ indicate that the solution is unstable and 
that there is a homoclinic orbit.
 The simple periodic eigenvalues 
are labeled by circles and the double points are labeled by crosses.
An example of a spectrum for a nearby semi-stable 
$N$-phase solution where the 
complex double point is split ${\mathcal O}(\eps)$ is given in Fig. 2(a).

Explicit formulas for the $N$-phase solutions, 
$\Theta(\theta_1,...,\theta_N)$,
are obtained in terms of the simple spectrum.
The phases evolve according to 
$\theta_j = \kappa_j x + \Omega_j t + \theta_j^{0}$,
$\kappa_j = 2\pi n_j/L$,
where $\kappa_j$ and $\Omega_j$ are determined 
by $\lambda_j^s$ (since the 
spectrum is invariant $\kappa_j$ and $\Omega_j$ are 
constants).
For a given $N$-phase solution, the isospectral set (all NLS solutions
with the same spectrum) comprises an $N$-dimensional
torus characterized by the phases $\theta_j$. If the spectrum 
contains complex double points, then
the $N$-phase solution may be unstable. The instabilities  
correspond to orbits homoclinic to the $N$-phase torus.

\noindent {\bf JONSWAP data and the proximity to homoclinic solutions 
of the NLS}
In the numerical simulations the NLS equation is integrated
using a pseudo-spectral scheme with $256$ Fourier modes in
space and a fourth order Runge-Kutta discretization in
time ($\Delta t = 10^{-3}$).
The nonlinear mode  content of
the data is numerically computed using the direct 
spectral transform described above, i.e. the 
system of ODEs \rf{lax1} is numerically solved to obtain
the discriminant $\Delta$.
The zeros of $\Delta \pm 2$ are then determined with a root solver 
based on Muller's method\cite{ref14}. 
The spectrum is computed with an accuracy of ${\mathcal O}(10^{-6})$,
whereas the spectral quantities 
we are interested in range from 
${\mathcal O}(10^{-2})$ to ${\mathcal O}(10^{-1})$.

Complex double points typically  split 
under perturbation into two simple points,
$\lambda_{\pm}$, thus opening a gap in the band of
spectrum (see Fig. 2(a)). 
We denote the distance between these two simple points by
$\delta(\lambda_+,\lambda_-) = |\lambda_+ - \lambda_-|$ and refer to it
as the splitting distance. We use $\delta$ to measure the 
proximity in the spectral plane to  homoclinic data, 
i.e. to complex double points  and their corresponding instabilities.
Since the NLS spectrum  is 
symmetric with respect to the real axis and real double points
 correspond to inactive modes, in subsequent plots
only the spectrum in
the upper half complex $\lambda$-plane will be displayed.

We begin by determining the spectrum of JONSWAP initial data given by
\rf{uinit} for various combinations of $\alpha = .008, .012, .016, .02$,
and $\gamma = 1, 2, 4, 6, 8$.
For each such pair $(\gamma, \alpha )$, we performed
fifty  simulations, each  with a different set of randomly
generated phases.
As expected, the basic spectral configuration
 and the number of excited modes depended on
the energy and the enhancement coefficient  $\alpha$ and
$\gamma$. However, the extent of the dependence of the spectrum 
upon the phases in the initial data  was surprising.

As a typical example of the results,
Figs. 2(a) and 3(a)  show  the numerically computed nonlinear spectrum of
JONSWAP initial data when $\gamma = 4$ and $\alpha = .016$ for
two different realizations of the random phases.
\begin{figure}[htb]
\centerline{\psfig{figure=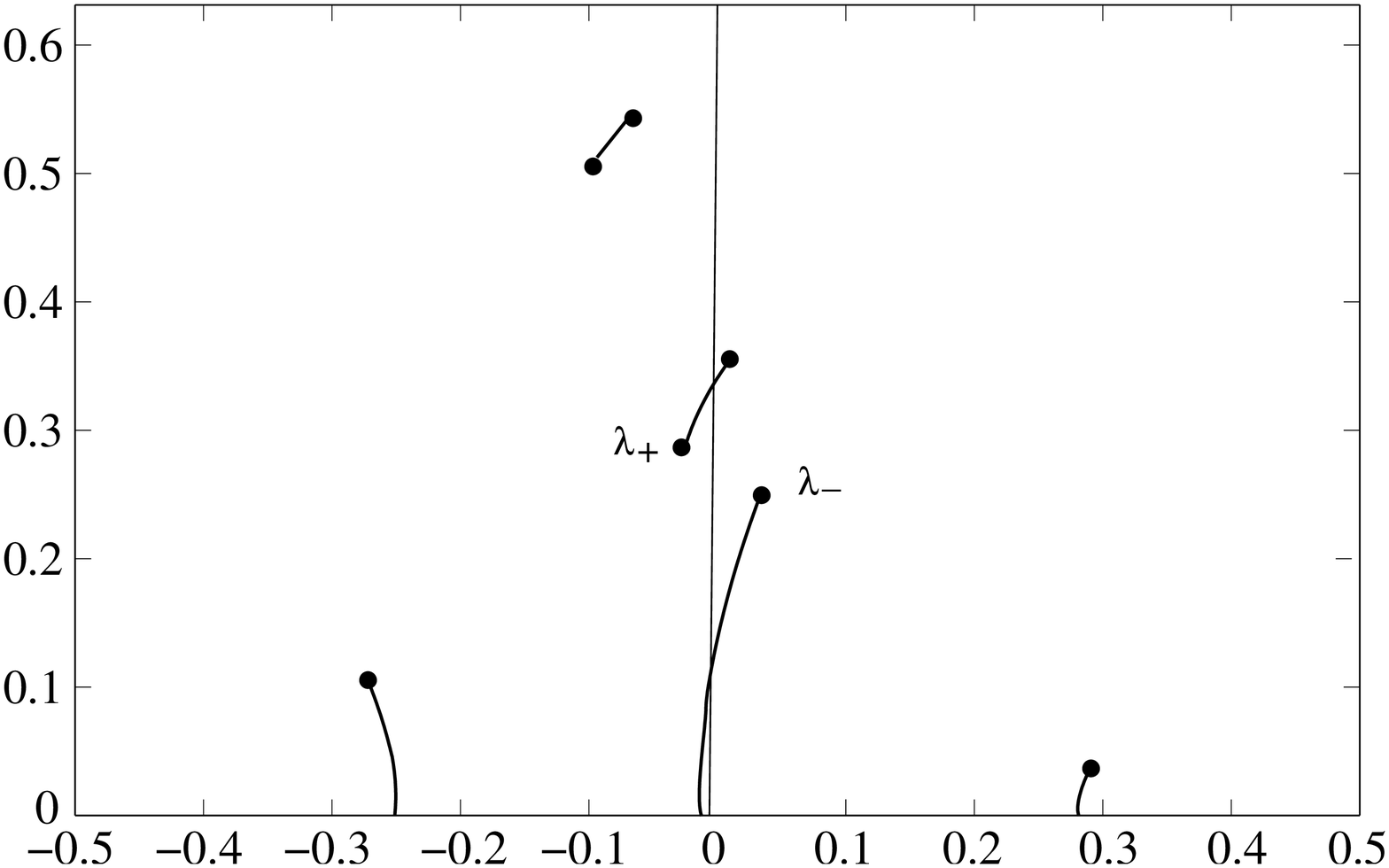,width=2.in,height=1.48in}
\hspace{24pt}\psfig{figure=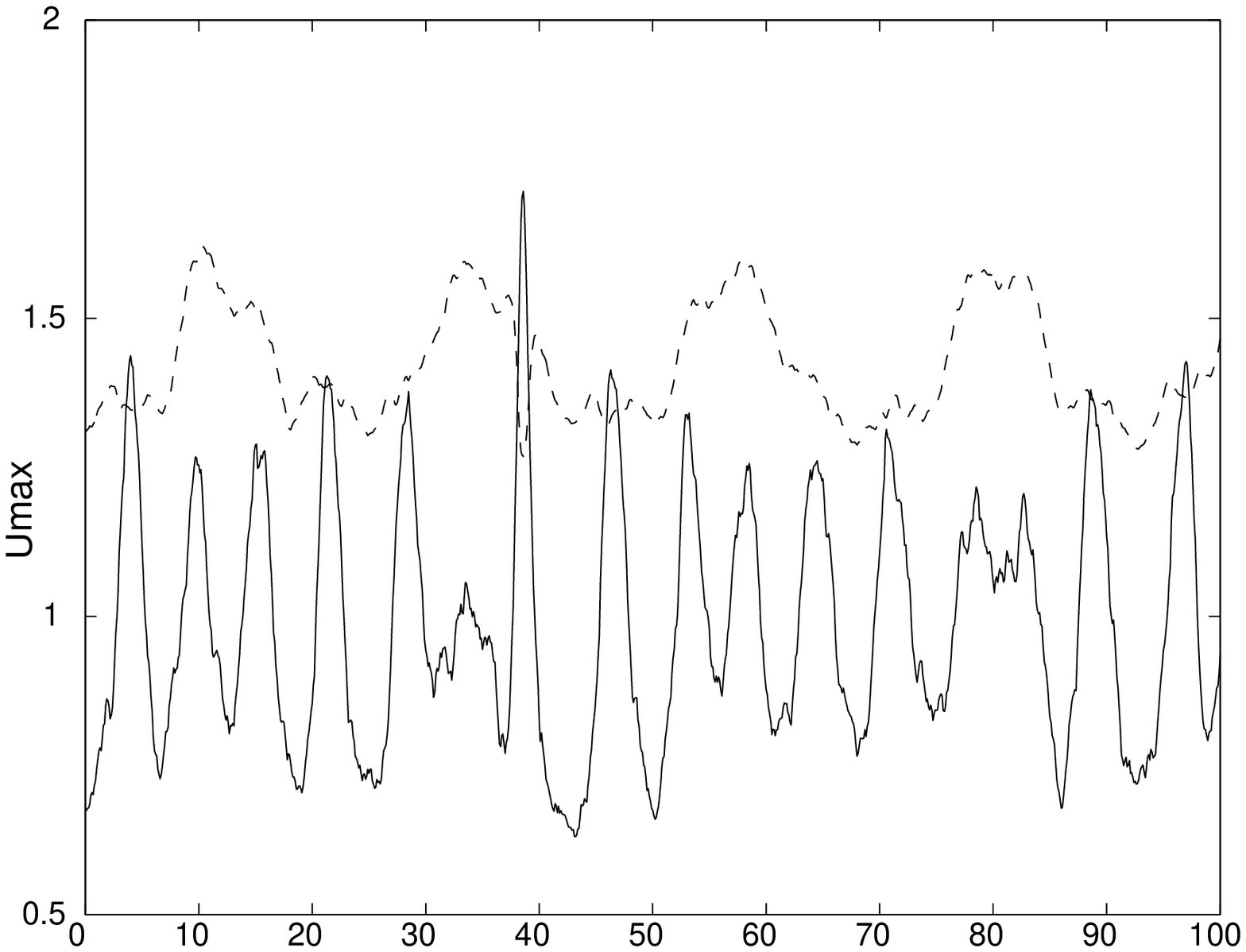,width=2.in}}
\centerline{(a)\hspace{150pt} (b)}
\caption{(a) Nonlinear spectrum and (b) evolution of $U_{max}$ 
for  JONSWAP data ($\gamma = 4$ and $\alpha = .016$) that is near 
homoclinic data. Dashed curve corresponds to 2.2$H_s$.}
\end{figure}
We find that JONSWAP data correspond to 
``semi-stable'' $N$-phase solutions, i.e. we 
interpret the data  as perturbations 
of $N$-phase  solutions with one or more unstable 
modes 
(compare Fig. 2(a)  with the spectrum
of an unstable $N$-phase solution in Fig. 1).
In Fig. 2(a) the splitting distance $\delta(\lambda_+,\lambda_-) \approx .07$,
while in 
Fig. 3(a)  $\delta(\lambda_+,\lambda_-) \approx .2$.
\begin{figure}[htb]
\centerline{\psfig{figure=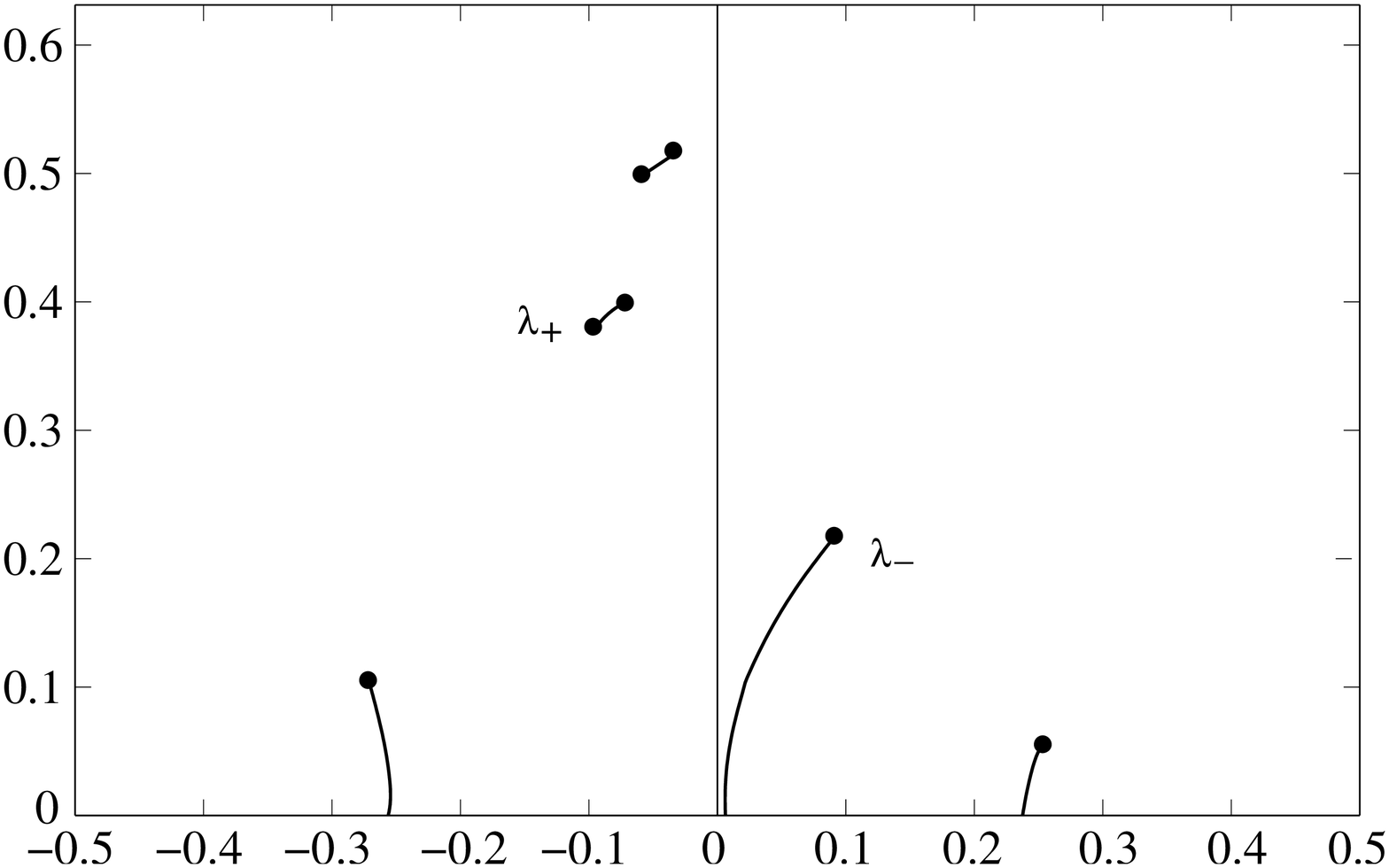,width=2.in,height=1.48in}
\hspace{24pt}\psfig{figure=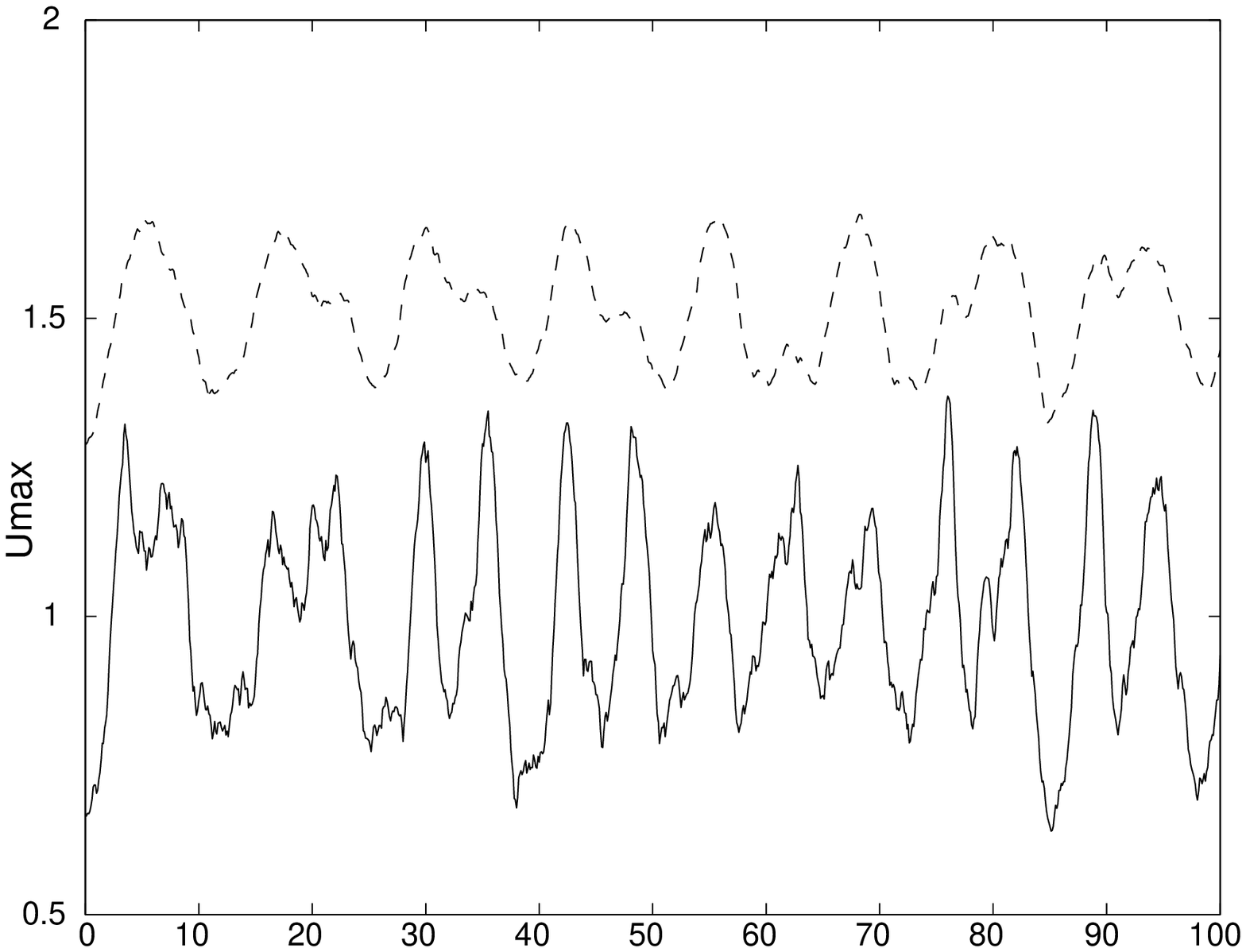,width=2.in}}
\centerline{(a)\hspace{150pt} (b)}
\caption{(a) Nonlinear spectrum and (b) evolution of $U_{max}$ for  
JONSWAP data ($\gamma = 4$ and $\alpha = .016$) that is 
far from homoclinic data. Dashed curve corresponds to 2.2$H_s$.}
\end{figure}
Thus the  JONSWAP data  can be quite ``near'' homoclinic data
as in Fig. 2(a)  or ``far'' from  homoclinic data as in Fig. 3(a),
depending on the values of the phases $\phi_n$ in the initial data.
For all the examined values of $\alpha$ and $\gamma$
we find that, when  $\alpha$ and $\gamma$ are fixed, as the phases in the 
JONSWAP data vary, the spectral  distance $\delta$
of typical JONSWAP data from homoclinic data varies.

Most importantly, irrespective of the values of the JONSWAP parameters 
$\alpha$ and $\gamma$,  in simulations of the NLS equation \rf{NLS}
we find that 
extreme waves develop 
for JONSWAP initial data that is ``near'' NLS homoclinic data,
whereas the JONSWAP data that is ``far'' from NLS homoclinic data 
typically does not generate extreme waves. 
Figs. 2(b) and 3(b) show the corresponding 
evolution of the maximum surface elevation,
$U_{max}$, obtained with the NLS equation.
$U_{max}$ is given by the solid curve and
as a reference, $2.2 H_S$ (the threshold for a rogue wave) 
is given by the dashed curve.
$H_S$ is the significant wave height and 
is calculated as four times the standard deviation of the 
wave amplitude.
Figure 2(b) shows that when the nonlinear spectrum 
is near homoclinic data,
$U_{max}$ exceeds $2.2 H_S$ 
(a rogue wave develops at about $t = 40$).
Figure 3(b) shows that when 
the nonlinear  spectrum  is far from 
homoclinic data,
$U_{max}$ is significantly below 
$2.2 H_S$ and a rogue wave  does not develop.
In this way, we correlate the occurrence of rogue waves 
characterized by JONSWAP
spectrum with the proximity to homoclinic solutions of the NLS equation.

The results of hundreds of simulations of the NLS equation
consistently show that 
proximity to homoclinic data is a crucial indicator
of rogue wave events.
For example, Fig. 4 shows the synthesis of 200 random simulations of the 
NLS equation for JONSWAP initial data for different $(\gamma,\alpha)$ pairs
(with $\gamma =  2, 4, 6, 8,$ and $\alpha =  .012, .016$).
\begin{figure}[htb]
\centerline{\psfig{figure=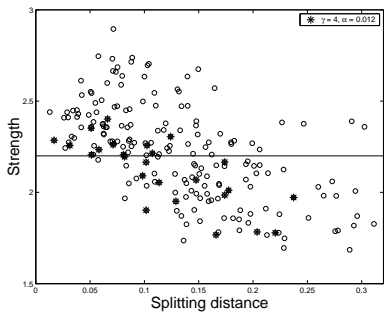,width=2.in}}
\caption{Strength of $U_{max}/H_s$ vs. the splitting distance 
$\delta(\lambda_+,\lambda_-)$.}
\label{jons4}
\end{figure}
For each such pair $(\gamma,\alpha)$, we performed
25 simulations, each  with a different set of randomly
generated phases. 
Each circle represents the 
strength of the maximum wave  ($U_{max}/H_S$) attained during one simulation
as a function of the splitting distance 
$\delta(\lambda_+,\lambda_-)$. The results for the particular pair 
$(\gamma = 4, \alpha = 0.012)$ is represented with an asterisk.
A horizontal line at 
$U_{max}/H_S = 2.2$ indicates the reference 
strength for rogue wave formation. 
We identify two critical values $\delta_1 = 0.08$ and $\delta_2 = 0.22$
that clearly show that (a) if $\delta < \delta_1$ (near homoclinic data)
rogue waves will occur;
(b) if $\delta_1 < \delta < \delta_2$, the  likelihood
of obtaining rogue waves decreases as $\delta$ increases and,
(c) if $\delta > \delta_2$  the likelihood of a rogue wave occurring 
is extremely small.

This behavior is robust. 
As  $\alpha$ and $\gamma$ are varied, the strength of the maximum wave
and the occurrence of rogue waves
are well predicted by the proximity to homoclinic solutions.
The individual plots of the strength vs. $\delta$
for particular pairs $(\gamma, \alpha)$
are qualitatively the same as in Fig. 4 as can be seen by the 
highlighted case $(\gamma = 4, \alpha = 0.012)$.
These results give strong evidence 
of the relevance of homoclinic solutions of the NLS equation
in investigating rogue wave phenomena for more 
realistic oceanic conditions and 
identifies the nonlinear spectral decomposition 
as a simple diagnostic tool
for predicting the occurrence and strength of rogue waves.
Finally we remark that the nonlinear spectral analysis
can be implemented for other general
data (non-JONSWAP) in order to predict the occurrence of rogue waves. 

This work was partially supported by NSF
Grant No. NSF-DMS0204714.

\end{document}